\documentclass[reprint,aps,amsmath,amssymb,amsfonts,showkeys]{revtex4-2}
\usepackage{graphicx}
\usepackage{dcolumn}
\usepackage{bm}
\usepackage{txfonts}
\usepackage{multirow}
\usepackage{color}
\newcommand{\authall}[1]{{}}

\usepackage{ulem}%use \sout{}

\begin{document}

\title{Anomalous scaling law for thermoelectric transport of 2D-confined electrons in an organic molecular system}

\author{Naoki~Kouda}
\author{Kyohei~Eguchi}
\author{Ryuji~Okazaki}
\email{okazaki@rs.tus.ac.jp}
\author{Masafumi~Tamura}
\affiliation{Department of Physics, Faculty of Science and Technology, Tokyo University of Science, Noda 278-8510, Japan}

\begin{abstract}

Confined electrons in low dimensions host desirable material functions for 
downscaled electronics as well as advanced energy technologies.
Thermoelectricity is a most fascinating example, 
since the dimensionality modifies the electron density of states dramatically, 
leading to enhanced thermopower
as 
experimentally 
examined in artificial two-dimensional (2D) structures.
However, it is 
still 
an open question
whether such an enhanced thermopower
in low dimensions 
is realized in layered materials with strong 2D characters such as cuprates.
Here, we report unusual enhancement of the thermopower in the layered organic compound $\alpha$-(BEDT-TTF)$_2$I$_3$,
where BEDT-TTF stands for bis(ethylenedithio)-tetrathiafulvalene.
We find that 
the slope in the Jonker plot (thermopower $S$ vs. logarithm of electrical conductivity $\log\sigma$) for $\alpha$-(BEDT-TTF)$_2$I$_3$ 
is significantly larger than that of conventional semiconductors.
Moreover, the large slope is also seen in the related layered salt,
demonstrating 
the impact of the 2D-confined carriers in the layered organics
on thermoelectricity.

\end{abstract}

\maketitle

\section{introduction}

Thermoelectricity, 
a fundamental property of solids to generate the electric field ${\bm E}$ under the temperature gradient $\nabla T$
with the proportional coefficient $S$ known as the thermopower or the Seebeck coefficient
as ${\bm E} = S\nabla T$ in an open circuit, 
offers a simple solid-state technology for the direct heat-to-electricity conversion,
yet
it is a very challenging
issue to
establish the guiding principles 
for the high-performance thermoelectrics \cite{Snyder2008,He2017}.
From a fundamental point of view,
the semiclassical Boltzmann approach gives an approximate 
formula of the thermopower for a degenerate electron gas,
which is well known as the Mott relation,
\begin{align}
S = \frac{\pi^2}{3}\frac{k_{\rm B}}{q}k_{\rm B}T\left.\frac{d\ln\sigma(\varepsilon)}{d\varepsilon}\right|_{\varepsilon=\mu},
\end{align}
where $k_{\rm B}$ is the Boltzmann constant, $q$ is the carrier charge, 
$\sigma$ is the electrical conductivity, and $\mu$ is the chemical potential \cite{Behniabook},
signifying a close link to the energy dependence of the conductivity.
Indeed, 
this relation underlies as a basal guideline for various schemes
such as the band structure \cite{Pei2011} 
and the mobility \cite{Sun2015} engineering,
in which the microscopic parameters in the conductivity formula
such as the density of states (DOS) and the 
relaxation time are successfully controlled to increase the thermopower.

Among the various concepts based on the Mott relation,
the reduced dimensionality is a straightforward and intriguing way 
as to adopt a step-like singularity in the DOS near the band edge.
If the electron chemical potential is close to the edge, 
as in the case of a narrow-gap semiconductor,
the energy derivative of the DOS is expected to diverge 
so as to afford extraordinarily large thermopower \cite{Hicks1993}.
This theoretical proposal has motivated well-conceived transport measurements 
on the artificial systems such as
the one-dimensional (1D) nanowires \cite{Hochbaum2008,Boukai2008}
and the two-dimensional (2D) superlattices \cite{Venkatasubramanian2001},
leading to the experimental demonstration of the improved 
dimensionless figure of merit, 
$ZT=S^2\sigma T/\kappa$, where 
$\kappa$ is the thermal conductivity,
although
these observations seem to come from the phonon effect \cite{Hochbaum2008,Boukai2008,Venkatasubramanian2000}
rather than the proposed DOS modification.
On the other hand, 
Ohta \textit{et al.} presented unusually large thermopower emerged from 
the 2D electron gas (2DEG) in the oxide superlattice \cite{Ohta2007},
indicating a 2D quantum confinement to vary the DOS.
Moreover, such a 2DEG has also been realized at the surface of the three-dimensional (3D) compounds
incorporated into the field-effect-transistor structure,
in which a systematic evolution of the thermopower of the 2D-confined carriers is 
achieved by the gate voltage tuning \cite{Ohta2012,Shimizu2016}.

%:figcr
\begin{figure*}[t!]
\begin{center}
\includegraphics[width=14cm]{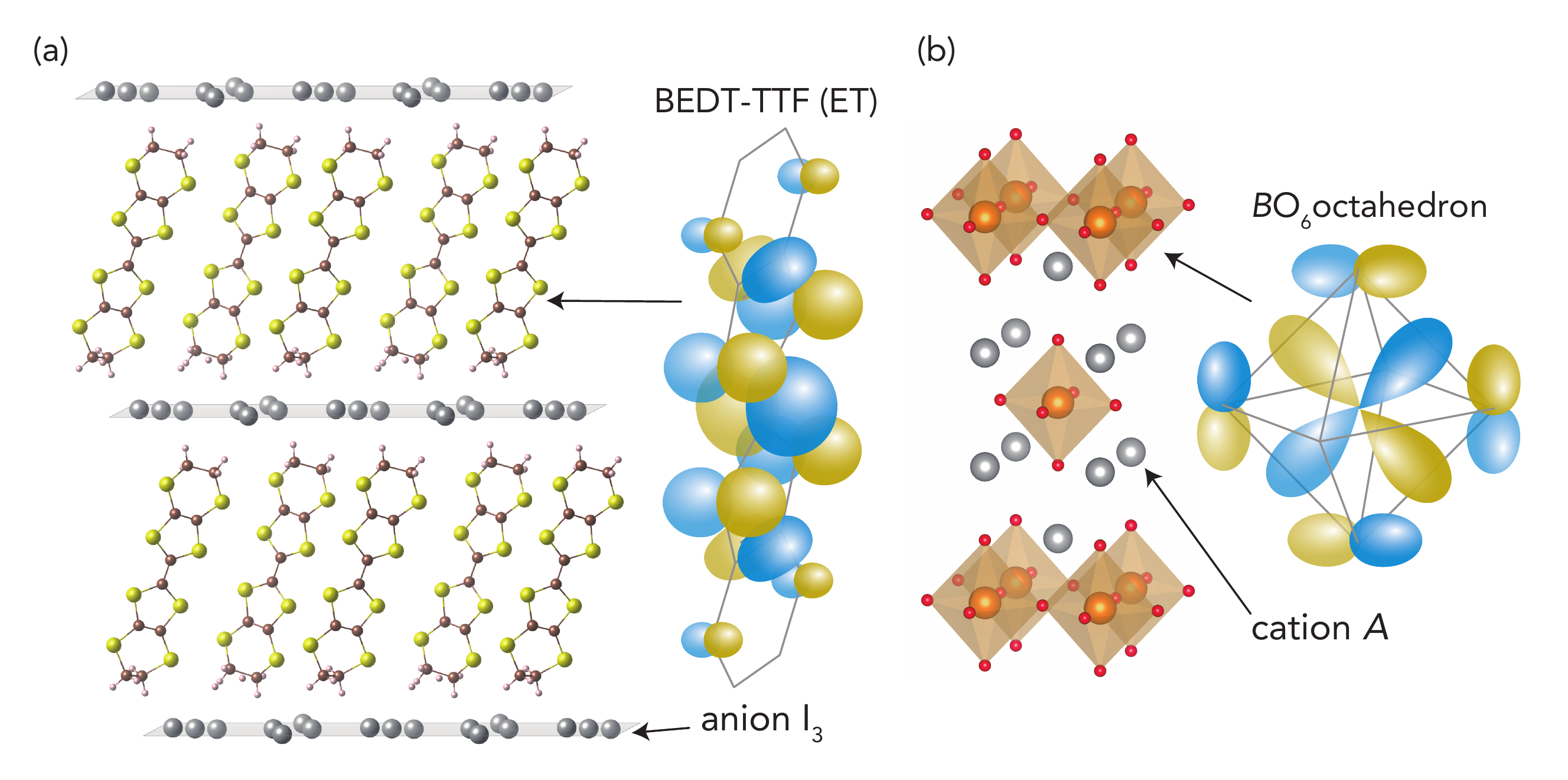}
\caption{
Schematic crystal structures of layered materials.
(a) 
Layered organic salt $\alpha$-(ET)$_2$I$_3$ consisting of planar BEDT-TTF (ET) molecules.
The highest occupied molecular orbital (HOMO) of ET molecule is schematically drawn.
The HOMO is spread along the direction normal to the molecular plane,
leading to $\pi$-stacked conducting layer with strong two-dimensional character.
(b)
Layered perovskite oxides $A_2B$O$_4$ consisting of three-dimensional $B$O$_6$ octahedra.
Schematic $t_{2g}$ orbital of $B$O$_6$ octahedron is shown.
}
\end{center}
\end{figure*}

A key question subsequently arises:
does such a drastic modification in DOS enhance thermopower in a bulk material with low dimensionality?
Many of remarkable physical phenomena have been found as a result of the low-dimensional structures.
Here, we focus on the charge transfer organic salt $\alpha$-(ET)$_2$I$_3$
[ET being bis(ethylenedithio)-tetrathiafulvalene (BEDT-TTF)],
in which the ET and the I$_3$ anion layers are alternatingly stacked to form 
the 2D layered crystal structure as illustrated in Fig. 1(a) \cite{Bender1984}.
This material exhibits a charge order transition at $T_{\rm CO}=136$~K \cite{Takano2001,Kakiuchi2007}, 
which
is driven by the inter-site Coulomb repulsion \cite{Seo2006}.
The two dimensionality in the charge order phase below $T_{\rm CO}$ is clearly 
evidenced by 
the anisotropy in the resistivity \cite{Ivek2017} as well as 
an occurrence of a Kosterlitz-Thouless transition at 
$T_{\rm KT}\approx35$~K \cite{Uji2013}. 
In this study, we performed the electrical conductivity $\sigma$ and the thermopower $S$ 
measurements on $\alpha$-(ET)$_2$I$_3$ single crystals with a systematic evaluation of the sample 
dependence.
The thermopower in the charge order phase is unusually large and incompatible with the conventional band picture,
but is well scaled in the $S$-$\log \sigma$ plot,
which is strikingly similar to that in the 2D-confined electrons realized in the oxide superlattices.

\section{experiments}

Single crystals of $\alpha$-(ET)$_2$I$_3$ were 
prepared by an electrochemical method.
The crystal orientation was determined from the polarized infrared reflectivity spectra measured 
by using a Fourier transform infrared spectrometer \cite{Meneghetti1986}.
The resistivity and the thermopower were simultaneously measured by using a conventional four-probe method and 
a steady-state method, respectively \cite{Takahashi2016,Yamanaka2022}. 
For the thermopower measurement, 
a manganin-constantan differential thermocouple was attached to the sample by using a carbon paste
and 
the temperature gradient ($|\nabla T| \approx$~0.5 K/mm) was applied by using a resistive heater.
The thermoelectric voltage from the wire leads was subtracted. 
The rate of temperature change is lower than 0.3~K/min to prevent the damage to the sample.

%:fig1
\begin{figure*}[t!]
\begin{center}
\includegraphics[width=18cm]{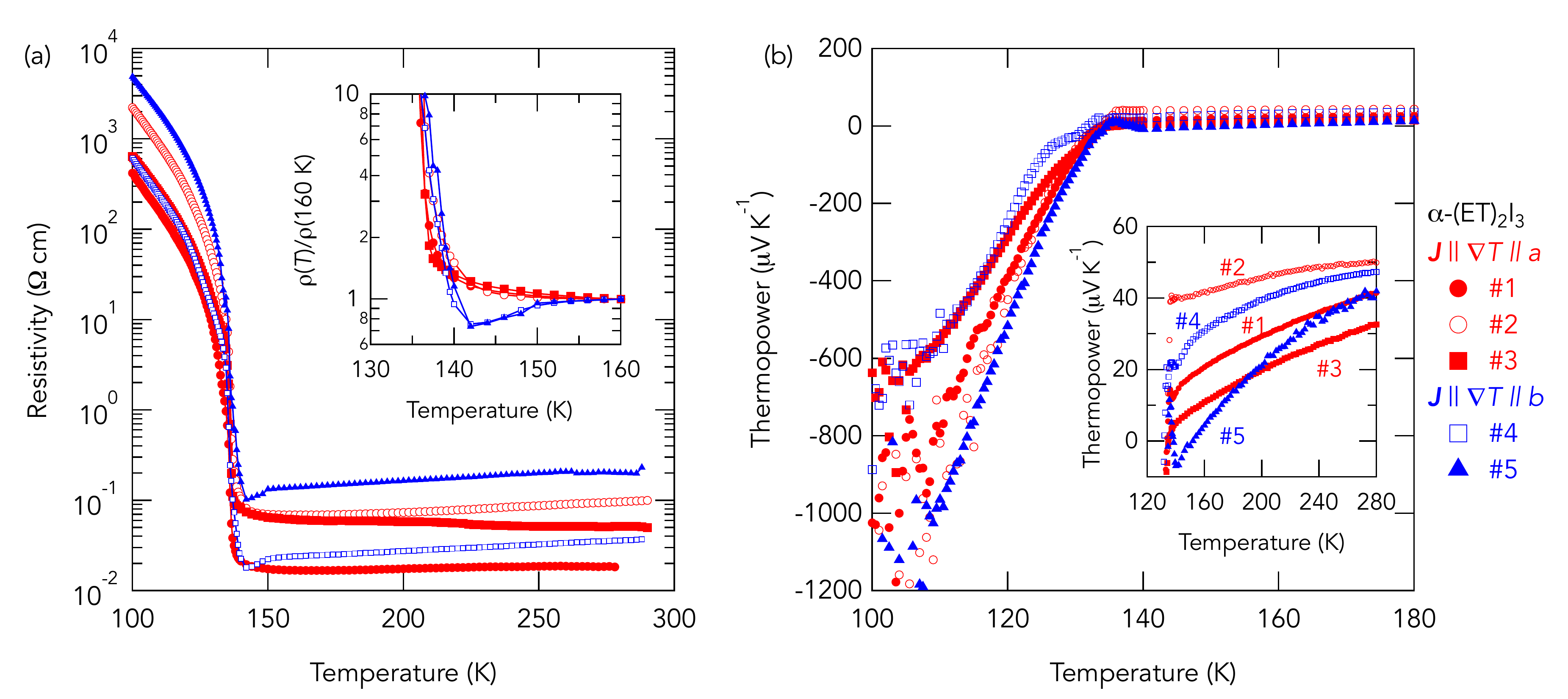}
\caption{
Transport properties of $\alpha$-(ET)$_2$I$_3$ single crystals.
(a), 
Temperature dependence of the resistivity $\rho$,
and 
(b),
the thermopower $S$ for five single crystals.
The direction of the electrical current density ${\bm J}$ and 
the temperature gradient $\nabla T$
is 
${\bm J}~||~\nabla T~||~a$ for the samples $\#1, \#2, \#3$ (red) and 
${\bm J}~||~\nabla T~||~b$ for the samples $\#4, \#5$ (blue).
The thermopower data below 110 K is fluctuating owing to high resistance at low temperatures.
Insets: 
(a) The normalized resistivity near the charge order transition temperature $T_{\rm CO}=136$~K.
(b) The thermopower in the high-temperature regime.
}
\end{center}
\end{figure*}

\section{results and discussion}

\subsection{Resistivity}

%:results

Figure 2(a) depicts the temperature dependence of the electrical resistivity $\rho$ of 
$\alpha$-(ET)$_2$I$_3$ single crystals. 
All the samples exhibit the metal-insulator transition at $T_{\rm CO}=136$~K owing to the 
charge order.
On the other hand, 
one may find the significant sample dependence in the magnitude of the resistivity shown in Fig. 2(a).
Although there is an inevitable ambiguity of the the sample size and the current path in the resistivity measurement
in general,
this sample dependence may also be intrinsic 
as observed in the thermopower.
In the inset of Fig. 2(a),
we plot the temperature dependence of the normalized resistivity by the value at $T=160$~K.
The in-plane anisotropy between the resistivity data measured for ${\bm J}~||~a$ ($\rho_{aa}$) and 
for ${\bm J}~||~b$ ($\rho_{bb}$) is clearly seen just above the transition temperature $T_{\rm CO}$:
while $\rho_{aa}$ exhibits a gradual decrease on heating, 
$\rho_{bb}$ shows a minimum structure near $T=145$~K,
which has also been observed in the earlier study \cite{Ivek2017}.
This in-plane resistive anisotropy may originate from the charge-disproportionation fluctuations 
existing even in the high-temperature phase \cite{Yue2010}:
the stripe-type charge order along the $b$ axis \cite{Takahashi2006,Katano2015}
may induce 
a smooth charge flow along the same direction
but form strong potential barriers 
along the $a$-axis direction 
to suppress the conduction.

\subsection{Thermopower}

Figures 2(b) displays the temperature dependence of the thermopower $S$.
The overall behavior of the thermopower is 
similar to earlier results that reported this material for the first time \cite{Bender1984}:
the thermopower is positive and relatively small value as expected in the metallic state,
while it changes its sign and shows the large absolute value in the insulating phase.
This behavior is also qualitatively consistent with the temperature dependence of the Hall coefficient \cite{Ivek2017}.
On the other hand, the detailed nature of the thermopower has been not discussed in the first report \cite{Bender1984},
in which the thermopower was given as evidence for the metal-insulator transition at $T_{\rm CO}$.
We also note that the thermopower of $\alpha$-(ET)$_2$I$_3$ has  
been intensively studied to investigate the exotic Dirac-like electronic states driven by pressure \cite{Konoike2013,Monteverde2013}.

The observed sample dependence of the transport properties may originate from the disorder effect,
which is suggested to be intrinsic in this material.
This is experimentally evidenced by the relaxor-type dielectric response \cite{Lunkenheimer2015,Ivek2017} 
in sharp contrast to the 1D charge order salts \cite{Nad2000} and 
the negative magnetoresistance in the charge order phase possibly due to the weak localization \cite{Ivek2017}.  
Although it requires detailed future study, the origin of disorders may stem from the anions layer \cite{Ivek2017}:
the I$_3^-$ anions are still chemically reactive as an oxidant in the crystal.
The transition temperature $T_{\rm CO}$ is, nevertheless, little affected, showing
the cohesive properties in the conduction layers are substantially retained.
Such a disorder effect, however, should affect on the transport properties seriously \cite{Alemany2012}.
In addition, 
the high-temperature phase of $\alpha$-(ET)$_2$I$_3$ above $T_{\rm CO}$ is semimetallic, consisting
of small number of electrons and holes ($n_{\rm e}\approx n_{\rm h}\approx 10^{18}$~cm$^{-3}$)
with high mobility ($\mu_{\rm e}\approx \mu_{\rm h}\approx 10^{2}$~cm$^{2}$V$^{-1}$s$^{-1}$),
where $n_{\rm e(h)}$ and $\mu_{\rm e(h)}$ are the carrier density and the mobility of the electrons (holes), respectively \cite{Ivek2017}.
In this case,
the thermopower is 
weighted by the conductivity of each carrier as
$
S = (\sigma_{\rm e}S_{\rm e}+\sigma_{\rm h}S_{\rm h})/(\sigma_{\rm e}+\sigma_{\rm h})
$,
where 
$\sigma_{\rm e(h)}=en_{\rm e(h)}\mu_{\rm e(h)}$ and $S_{\rm e(h)}$ are the 
conductivity and the thermopower of electrons (holes), respectively.
Hence, the thermopower should  be quite sensitive to the 
balance of the contributions from electrons and holes;
the predominant carrier in the semimetallic state above $T_{\rm CO}$ is hole as also indicated from the Hall coefficient \cite{Ivek2017}, 
but the aforementioned disorder may influence the delicate balanced carrier density and mobility,
resulting in the sample-dependent thermopower as seen in the inset of Fig. 2(b).

%:fig3
\begin{figure*}[t!]
\begin{center}
\includegraphics[width=18cm]{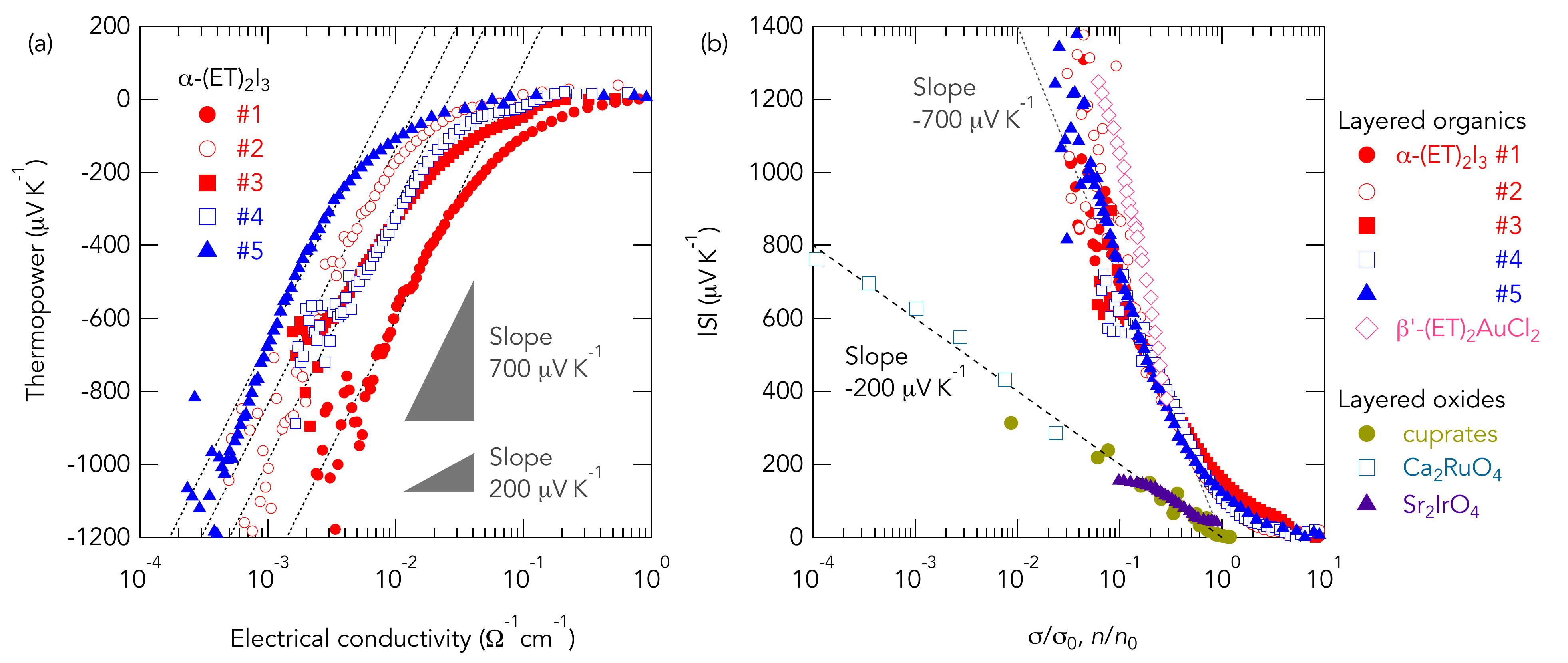}
\caption{
Jonker plot.
(a), 
The thermopower as a function of the conductivity in the single logarithmic plot 
for five single crystals of $\alpha$-(ET)$_2$I$_3$.
In conventional semiconductors, the slope of $200$~$\mu$V/K is expected.
On the other hand, 
unusual value of the slope of $700$~$\mu$V/K has been observed in 
the charge order phase of $\alpha$-(ET)$_2$I$_3$ in common.
The dotted lines are a guide to show the slope of $700$~$\mu$V/K for the low-temperature data.
(b), 
The Jonker plot for various layered materials including 
charge-order insulator $\alpha$-(ET)$_2$I$_3$ (present study),
dimer-Mott insulator $\beta'$-(ET)$_2$AuCl$_2$ \cite{Taniguchi2005,Kiyota2018},
cuprates \cite{Obertelli1992},
layered Mott insulator Ca$_2$RuO$_4$ \cite{Nishina2017}, 
and 
layered spin-orbit Mott insulator Sr$_2$IrO$_4$ \cite{Pallecchi2016}.
The vertical axis shows the magnitude of the thermopower.
The horizontal axis is the electrical conductivity $\sigma$ (or the carrier density $n$) normalized by the horizontal intercept $\sigma_0$ (or $n_0$) of the slope line for each compound.
The dashed and dotted lines represent the values of the slope of $-200$~$\mu$V/K and 
$-700$~$\mu$V/K, respectively.
}
\end{center}
\end{figure*}

\subsection{Jonker analysis}

To explore the universality among the strongly sample-dependent transport properties,
we employ the thermopower plotted as a function of the logarithm of the electrical conductivity $\sigma=\rho^{-1}$,
known as the Jonker plot \cite{Jonker1968}, 
in Fig. 3(a).
Quite interestingly, although the resistivity and the thermopower are sample-dependent as mentioned,
the slope of the $S$-$\log\sigma$ plot in the insulating phase seems to be nearly equal among the measured samples 
as guided by the dotted lines.
Note that,
in the semiconductors with the thermal activation energy gap of $E_{\rm g}$,
the electrical conductivity and the thermopower are expressed as
$\sigma=\sigma_0\exp(-E_{\rm g}/2k_{\rm B}T)$ and 
$S = (k_{\rm B}/q)(E_{\rm g}/2k_{\rm B}T)$, respectively 
($\sigma_0$ being a constant),
leading to a relationship between $S$ and $\ln\sigma$ as
\begin{align}
S = -\frac{k_{\rm B}}{q}(\ln\sigma -\ln\sigma_0),
\label{jonker}
\end{align}
and hence the magnitude of the slope becomes a universal value of $k_{\rm B}/|q|\approx 86$~$\mu$V/K 
($k_{\rm B}\ln10/|q|\approx 200$~$\mu$V/K),
which well holds in several conventional semiconductors \cite{Muta2008}.
On the other hand, the slope of the $S$-$\log\sigma$ plot
in the insulating phase of $\alpha$-(ET)$_2$I$_3$
is about $700$~$\mu$V/K in common, 
which is much larger than the conventional value of $200$~$\mu$V/K,
indicating an unusual mechanism to enhance the thermopower operating in $\alpha$-(ET)$_2$I$_3$.
Note that the intersection of the dotted line to $S=0$ in Fig. 3(a),
which corresponds to $\ln\sigma_0$ in Eq. (\ref{jonker}),
is sample-dependent.
This is consistent with the presence of the sample-dependent disorder,
since the impurity concentration is generally included in $\sigma_0$.
(The error of the sample size in determination of the resistivity could also be involved in this horizontal intercept.)

The origin of the enhanced thermopower in $\alpha$-(ET)$_2$I$_3$ is most crucial.
The two-carrier effect is excluded because it generally induces the cancellation of the thermopower.
Moreover, the temperature dependence of the mobility is also unlikely,
since temperature variations of the resistivity and the Hall coefficient are fairly scaled in the insulating phase \cite{Ivek2017}.
On the other hand, 
large slope in the Jonker plot has been reported in the 
2DEG realized in the oxide superlattice and discussed 
in terms of the modified DOS for the 2D-confined electrons:
while the series of the 3D doped SrTiO$_3$ bulk samples exhibits the conventional value of the slope of $200$~$\mu$V/K in the Jonker plot,
the 2D SrTiO$_3$ superlattice yields a large value of the slope of $1000$~$\mu$V/K \cite{Ohta2007}.
The remarkable resemblance to the present results evidently suggests 
a similar modification of the DOS acting for the correlated electrons confined in the 2D layers of $\alpha$-(ET)$_2$I$_3$.
Although the conventional band picture is not applicable to the present charge order system with strong electron correlation,
we speculate that the edge of the DOS, 
where the carriers are thermally excited, 
may show a steep change as a function of energy owing to the two dimensionality,
leading to the enhanced thermopower.
This effect can be enhanced by the in-plane anisotropy, 
or pseudo-1D character 
at the bottom of the upper band (electron carriers), 
as pointed out by the calculations \cite{Alemany2012}.

\subsection{Comparison with other 2D systems}

It is interesting to compare the present system with the other correlated layered systems.
Figure 3(b) depicts 
the Jonker plot for various layered materials including 
cuprates \cite{Obertelli1992},
layered Mott insulator Ca$_2$RuO$_4$ \cite{Nishina2017}, 
and 
layered spin-orbit Mott insulator Sr$_2$IrO$_4$ \cite{Pallecchi2016},
scaled with the normalized
electrical conductivity $\sigma$ (or the carrier density $n$) by the horizontal intercept $\sigma_0$ (or $n_0$) of the slope line for each compound.
The Jonker plot for the layered oxides shows a conventional slope value of $200$~$\mu$V/K.
Contrastingly, we find that 
the data of layered organic dimer-Mott insulator $\beta'$-(ET)$_2$AuCl$_2$, 
the resistivity and the thermopower of which are respectively extracted from Refs. \onlinecite{Taniguchi2005} and \onlinecite{Kiyota2018}, 
exhibits anomalously large slope similarly to $\alpha$-(ET)$_2$I$_3$.
This unique behavior in the layered organic salts
may stem from the highly anisotropic nature of the molecular orbitals [Fig. 1(a)]:
the $\pi$-lobes of the ET molecules spread along the in-plane directions strengthen the 2D nature,
while the typical layered oxides such as the layered perovskite $A_2B$O$_4$ are 
composed of rather 3D units of octahedra $B$O$_6$ [Fig. 1(b)].

The prominent two dimensionality in organics is clearly seen in the 
resistivity anisotropy $\rho_{\perp}/\rho_{\parallel}$ 
($\rho_{\perp}$ and $\rho_{\parallel}$ being the resistivity measured for the out-of-plane and the in-plane directions, respectively).
In the insulating phase of $\alpha$-(ET)$_2$I$_3$, it yields 
$\rho_{\perp}/\rho_{\parallel} \approx 10^{3}$ \cite{Ivek2017}.
This is significantly larger than that of layered oxide insulators.
For instance, while the anisotropy in the 2D correlated metal Sr$_2$RuO$_4$ is as large as $\rho_{\perp}/\rho_{\parallel}\approx10^3$
owing to the 2D cylindrical Fermi surface \cite{Yoshida1998},
it becomes almost unity 
$\rho_{\perp}/\rho_{\parallel}\approx1$
in the isovalent 2D Mott insulator Ca$_2$RuO$_4$ \cite{Nakamura2002}.
The anisotropy $\rho_{\perp}/\rho_{\parallel}$ also decreases near the Mott insulating phase of the cuprates
so that the system becomes less two dimensional \cite{Komiya2002}.
Thus, the present study serves a fascinating playground to examine the thermoelectricity 
of the 2D-confined correlated electrons in the bulk organic compounds,
opening up a particular route to apply the low-dimensional organic thermoelectrics.
It is noteworthy that the charge glass organic systems exhibit a large resistive anisotropy 
of $\rho_{\perp}/\rho_{\parallel} \approx 10^{6}$ at low temperatures \cite{Sato2020}.
Also, the 1D organic conductor exhibits a colossal thermopower \cite{Machida2016},
in which the low dimensionality should be essential.

\section{summary}

To summarize, 
we have systematically measured the electrical resistivity and the thermopower of $\alpha$-(ET)$_2$I$_3$ single crystals.
Although the transport properties are sample-dependent, 
the thermopower is well scaled as a function of the conductivity in the Jonker plot.
Importantly, 
the slope of the $S$-$\log\sigma$ plot in the insulating phase 
is unusually larger than that in conventional semiconductors, 
indicating 2D-confined correlated carriers to enhance the thermopower.

\subsection*{Acknowledgments}
We thank M. Ogoshi for experimental assistance of the infrared spectroscopy measurement.
This work was supported by JSPS KAKENHI Grants No. 18K13504.

%:ref
%%%%%%%%%%%%%%%%%%%%%%%%%%%%%%%%%%%%%%%%%%%%%%%%%%%%%%%%%%%%%%%%%%%%%%%%%%%%%%%%%
%\input{reference}
%\section{reference}

\end{document}